\documentclass[pra,showpacs,nobalancelastpage,onecolumn]{revtex4}

\usepackage{amssymb}

\usepackage{graphicx}
\usepackage{amsmath}

\begin{document}

\title{Continuous variable encoding by ponderomotive interaction}
\author{Stefano Pirandola, Stefano Mancini, David Vitali and Paolo Tombesi}

\affiliation{Dipartimento di Fisica, Universit\`a di Camerino, via
Madonna delle Carceri 9, I-62032 Camerino, Italy}

\date{\today}

\begin{abstract}
Recently it has been proposed to construct quantum error-correcting codes
that embed a finite-dimensional Hilbert space in the infinite-dimensional
Hilbert space of a system described by continuous quantum variables [D.
Gottesman \textit{et al.}, Phys. Rev. A \textbf{64}, 012310 (2001)]. The
main difficulty of this continuous variable encoding relies on the physical
generation of the quantum codewords. We show that ponderomotive interaction
suffices to this end. As a matter of fact, this kind of interaction between
a system and a meter causes a frequency change on the meter proportional to
the position quadrature of the system. Then, a phase measurement of the
meter leaves the system in an eigenstate of the stabilizer generators,
provided that system and meter's initial states are suitably prepared. Here
we show how to implement this interaction using trapped ions, and how the
encoding can be performed on their motional degrees of freedom. The
robustness of the codewords with respect to the various experimental
imperfections is then analyzed.
\end{abstract}

\pacs{03.67.Pp, 32.80.Lg, 03.65.Ta}

\maketitle

\section{Introduction}

Quantum information, as well as classical information, can be carried either
by discrete or continuous variable (CV) systems. The latter have attracted
an increasing interest during last years \cite{cvbook}. In particular
quantum error correction (QEC) techniques have been extended to this
framework in order to allow CV quantum computation \cite{Lloyd2,preskill}.
However, CV QEC presents some different aspects with respect to that
concerning qubits. In fact, while it is reasonable to encode a qubit in a
block of qubits using a QEC code protecting against a large error on a
single qubit of the block, this is no more true in CV systems. In such a
case, in fact, the most probable effect caused by decoherence is a small
diffusion in the position and momentum of all the particles. This is just
the kind of noise treated in \cite{preskill} where \textit{shift-resistant}
quantum codes for qudits are suitably extended to CV systems.
The main drawback of such codes is the preparation of the encoded states
which ideally are non-normalizable states and therefore can only be
approximated introducing an intrinsic error probability.
Travaglione and Milburn \cite{Travaglione0} have shown that it is
possible to generate such approximate states by performing a sequence of
operations similar to a quantum random walk algorithm \cite{Travaglione}.
More recently, we have proposed an all-optical scheme based on the
cross-Kerr interaction to realize such a code \cite{epl}. In this paper we
follow the suggestion of Ref.~\cite{preskill} and we study a \emph{
ponderomotive} interaction \cite{Giovannetti} to embed a qubit in a CV
quantum system.

The paper is
organized as follows. In Sec.~\ref{ideal} we rapidly review some elements
from Ref.~\cite{preskill} and we turn from an ideal situation to a more
realistic one. In Sec.~\ref{trap_scheme} the physical implementation of the
CV encoding scheme is proposed. Sec.~\ref{conclusion} is for conclusions.

\section{Encoded states\label{ideal}\label{encodedStateS}}

\begin{figure}[ptbh]
\vspace{-0.0cm}
\par
\begin{center}
\includegraphics[width=0.8\textwidth] {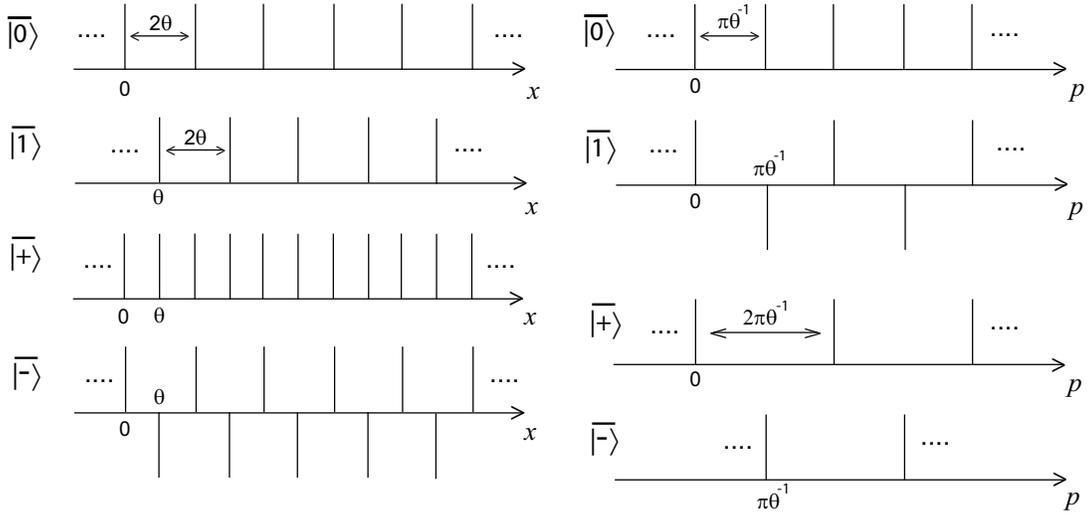}
\end{center}
\par
\vspace{-0.3cm}
\caption{Ideal encoded states $\overline{\left| 0\right\rangle },\overline{
\left| 1\right\rangle }$ ($\bar{Z}$ eigenstates) and $\overline{\left|
+\right\rangle },\overline{\left| -\right\rangle }$ ($\bar{X}$ eigenstates).
On the left the structure of the spatial wavefunctions is displayed while on
the right the structure of the momentum wavefunction is displayed. Each
spike is ideally a Dirac-delta function.}
\label{fig00}
\end{figure}

A single qubit living in a Hilbert space $\mathcal{H}$ with basis $\{\left|
0\right\rangle ,\,\left| 1\right\rangle \}$ can be encoded into a single
oscillator in such a way that the two resulting codewords $\overline{\left|
0\right\rangle },\overline{\left| 1\right\rangle }$ provide protection
against small diffusion errors in both position $x$ and momentum $p$ (the
quantum operators obey the commutation rule $[\hat{x},\hat{p}]=i$ so that $
x,p$ are dimensionless quantities). The two quantum codewords $\overline
{\left| 0\right\rangle },\overline{\left| 1\right\rangle }$ are the
simultaneous eigenstates, with eigenvalue $+1$, of the displacement operators
$\hat{D}_{x}(2\theta)=e^{-2i\theta\hat{p}}$, $\hat{D}_{p}(2\pi\theta^{-1})=e^{2i\pi\theta^{-1}\hat{x}}$, with $\theta \in
\mathbb{R}$, which are also the stabilizer generators of the code \cite{stabilizer}. These codewords are therefore invariant under the shifts $
x\rightarrow x-2\theta$ and $p\rightarrow p-2\pi\theta^{-1}$. Up to a
normalization factor they are given by
\begin{align}
\overline{\left| 0\right\rangle } & =\sum_{s=-\infty}^{+\infty}\left|
x=2\theta s\right\rangle =\sum_{s=-\infty}^{+\infty}\left| p=\pi\theta
^{-1}s\right\rangle ,  \label{ideal_0} \\
\overline{\left| 1\right\rangle } & =\sum_{s=-\infty}^{+\infty}\left|
x=2\theta s+\theta\right\rangle =\sum_{s=-\infty}^{+\infty}(-1)^{s}\left|
p=\pi\theta^{-1}s\right\rangle  =\hat{D}_{x}(\theta)\overline{\left| 0\right\rangle },  \label{ideal_1}
\end{align}
i.e. they are a coherent superposition of infinitely squeezed states
(position eigenstates and momentum eigenstates). Each of them is a
comb-state both in $x$ and in $p$ with equally spaced spikes ($2\theta$ in $
x $ and $\pi \theta^{-1}$ in $p$). The codewords $\overline{\left|
0\right\rangle },\overline{\left| 1\right\rangle }$ are also eigenstates of
the encoded bit-flip operator $\bar{Z}=\hat{D}_{p}(\pi\theta^{-1})$.
Equivalently one can also choose the codewords $\overline{\left|
\pm\right\rangle }=[\overline {\left| 0\right\rangle }\pm\overline{\left|
1\right\rangle }]/\sqrt{2}$ which are the eigenstates of the encoded
phase-flip operator $\bar{X}=\hat {D}_{x}(\theta)$ and are given by:
\begin{align}
\overline{\left| +\right\rangle } & =\sum_{s=-\infty}^{+\infty}\left|
x=\theta s\right\rangle =\sum_{s=-\infty}^{+\infty}\left| p=2\pi\theta
^{-1}s\right\rangle ,  \label{ideal_piu} \\
\overline{\left| -\right\rangle } &
=\sum_{s=-\infty}^{+\infty}(-1)^{s}\left| x=\theta s\right\rangle
=\sum_{s=-\infty}^{+\infty}\left|
p=2\pi\theta^{-1}s+\pi\theta^{-1}\right\rangle .  \label{ideal_meno}
\end{align}
Also these states are comb-like states both in $x$ and in $p$, with equally
spaced spikes ($\theta$ in $x$ and $2\pi\theta^{-1}$ in $p$). The four
codewords states are schematically displayed in Fig.~\ref{fig00}.

The recovery process is realized by measuring the stabilizer generators $
\hat{D}_{x}(2\theta),$ $\hat{D}_{p}(2\pi\theta^{-1})$. The measurement of
the $X$-generator $\hat{D}_{x}(2\theta)=\hat{p}($mod$
\pi\theta^{-1})$ reveals momentum shifts $\Delta p$ which are correctable if
$\left| \Delta p\right| <\pi\theta^{-1}/2$; in such a case the correction is
made by shifting $p$ so to become equal to the nearest multiple of $
\pi\theta^{-1}$. In the same way, the measurement of the $Z$-generator $\hat{
D}_{p}(2\pi\theta^{-1})=\hat{x}($mod$\theta)$
reveals position shifts which are correctable if $\left| \Delta x\right|
<\theta/2$; in such a case the correction is made by shifting $x$ so to
coincide with the nearest multiple of $\theta$.

Ref.~\cite{preskill} proposed the following recipe for the generation of the
codeword states.

\begin{enumerate}
\item  Preparation of a particle in the $p=0$ eigenstate (i.e. completely
delocalized in position).

\item  Coupling the particle to a meter (i.e. an oscillator, with ladder
operators $\hat{c}$, $\hat{c}^{\dagger }$) via the non linear interaction $
\hat{H}_{NL}=g\hat{c}^{\dagger }\hat{c}\hat{x}$. This interaction modifies
the frequency of the meter by $\Delta \omega =gx$ so that, at time $t=\pi
\theta ^{-1}g^{-1}$, the phase of the meter is shifted by $\Delta \phi =\pi
\theta ^{-1}x$.

\item  Reading out the phase of the meter $\Delta \phi $ at a time $t$, i.e.
measuring $\hat{x}($mod$2\theta )$. This measurement projects the initial
state into a superposition of equally spaced delta function $\delta
(x-2\theta s+\varepsilon )$ with $s=0,\pm 1,...$ and $\varepsilon \in
\mathbb{R}$.

\item  Applying a suitable transformation to obtain any desired encoded
qubit state $a\overline{\left| 0\right\rangle }+b\overline{\left|
1\right\rangle }$.
\end{enumerate}

Ideally the codewords are non-normalizable states infinitely squeezed both
in $x$ and $p$, but in practice one can only generate states with finite
squeezing, i.e. only approximate codewords: $\widetilde{\left|
0\right\rangle }\sim\overline{\left| 0\right\rangle },$ $\widetilde{\left|
1\right\rangle }\equiv\hat{D}_{x}(\theta)\widetilde{\left| 0\right\rangle }
\sim \overline{\left| 1\right\rangle },$ $\widetilde{\left| \pm\right\rangle
}\equiv\lbrack\widetilde{\left| 0\right\rangle }\pm\widetilde{\left|
1\right\rangle }]/\mathcal{N}_{\pm}\sim\overline{\left| \pm\right\rangle }$ (
$\mathcal{N}_{\pm}$ are normalization constants). For this reason, in order
to estimate the \textit{quality} of the encoding scheme,
together with the error
probability in the recovery process due to the occurrence of an
uncorrectable error, we have also to consider the \textit{intrinsic error
probability} due to the imperfections of the approximate codewords which can
lead to an error even in the presence of a correctable error. Here, we
propose a physical implementation of the ideal coding protocol of Ref.~\cite{preskill},
based on the interaction between an atom and a radiation mode.
In general, this scheme can be derived from the ideal one by replacing the
initial $p=0$ state with a finitely squeezed state, $\hat{H}_{NL}$ with a
ponderomotive interaction, and the phase measurement with a homodyne
measurement.

\section{Ponderomotive encoding\label{trap_scheme}}

\begin{figure}[ptbh]
\begin{center}
\includegraphics[width=0.5\textwidth] {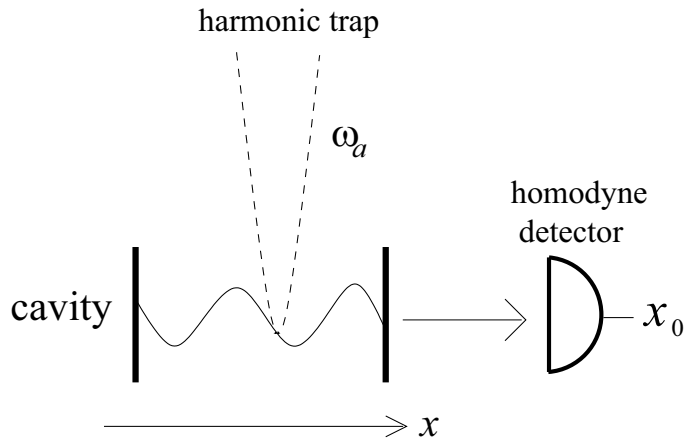}
\end{center}
\par
\vspace{-0.5cm}
\caption{\textit{Ponderomotive encoding}. An ion is trapped by a harmonic
potential within a high finesse cavity, where it interacts with a single
mode. After a suitable interaction time, a homodyne measurement of the
intracavity quadrature $\hat{X}=(\hat{c}+\hat{c}^{\dagger})/\sqrt{2}$ projects the
state of the ion onto an approximate comb-like state, i.e. the approximate
codeword $\widetilde{|0\rangle}$. A conditional displacement can then be
used to generate any correctable state $a\widetilde{|0\rangle}+b\widetilde
{|1\rangle}$ (see text).}
\label{fig02}
\end{figure}

Our proposal for a physical implementation of the CV encoding scheme is
schematically depicted in Fig.~\ref{fig02}.

We consider a single two-level ion (of mass $M$ and internal transition frequency $
\omega_{0}=2\pi c/\lambda_{0}$) trapped inside a high finesse cavity by a
harmonic potential, with trapping frequency $\omega_{a}$ along the direction
of the cavity axis $x$ (see Ref.~\cite{blatt} for experimental schemes of
this kind). We assume that the trapping potential in the other two
directions is steep enough to freeze the motion along $y$ and $z$. The ion
interacts with a single cavity mode with annihilation operator $\hat{c}$ and
frequency $\omega_{c}=2\pi c/\lambda_{c}=ck_{c}$. If the harmonic potential
minimum is set halfway in between a node and an antinode of the stationary
field of the cavity mode, the ion-cavity Hamiltonian can be written as
(assuming, as usual in the optical domain, the rotating wave and dipole
approximation)
\begin{align}
\hat{H} =\hbar\omega_{a}\hat{a}^{\dagger}\hat{a}+\hbar\omega_{c}\hat{c}
^{\dagger}\hat{c}+\hbar\omega_{0}\hat{\sigma}_{z}
+\hbar g_{0}\left( \hat{\sigma}^{\dagger}\hat{c}+\hat{c}^{\dagger }
\hat{\sigma}\right) \cos(k_{c}\hat{x}+\pi/4).  \label{trapped_ham0}
\end{align}
In this Hamiltonian $\hat{\sigma}_{z}=(\hat{\sigma}^{\dagger}\hat{\sigma}-
\hat{\sigma}\hat{\sigma}^{\dagger})/2$, $\hat{\sigma},\hat{\sigma}^{\dagger}$
are the atomic spin-$1/2$ operators associated with the two internal levels
whose transition is quasi-resonant with the optical cavity mode,
$\hat{x}=\sqrt{\hbar/(2M\omega_a)}(\hat{a}+\hat{a}^{\dagger})$
is the atomic center-of-mass (CM) position operator, $\hat{c},\hat{c}^{\dagger}$
are the cavity mode annihilation and creation operators, and $g_{0}$ is the
atom-field coupling constant.

A ponderomotive interaction between the atom and the cavity mode is obtained
in the dispersive limit in which the cavity mode is highly (red) detuned
from the atomic transition. In this limit, the upper atomic level can be
adiabatically eliminated and also the spontaneous emission from it can be
neglected. In such a condition the atom always remains in its ground state
and the resulting ponderomotive Hamiltonian
(in interaction picture with respect to
$\hat{H}_0=\hbar\omega_{c}\hat{c}^{\dagger}\hat{c}$), is
\begin{align}
\hat{H} =\hbar\omega_{a}\hat{a}^{\dagger}\hat{a}+\hbar
\frac{g_0^2}{\delta}\hat{c}^{\dagger}\hat{c}
\cos^2(k_{c}\hat{x}+\pi/4),  \label{trapped_ham1}
\end{align}
where $\delta=\omega_0-\omega_c$.
We then make a second assumption, the Lamb-Dicke limit, which amounts to
assume that the ion CM is well localized with respect to a wavelength of the
cavity mode $\lambda_{c}$. This assumption allows us to linearize the
resulting optical potential in the far-off resonance regime. We thus end up
with the simple Hamiltonian \cite{stefano}
\begin{equation}
\hat{H}=\hbar\omega_{a}\hat{a}^{\dagger}\hat{a}
-\hbar \frac{g_0^2}{2\delta}\hat{c}^{\dagger}\hat{c}
-\hbar g\hat{c}^{\dagger}\hat{c}\left( \hat{a}+\hat
{a}^{\dagger}\right) \,,  \label{trapped_ham}
\end{equation}
where $g=(g_{0}^{2}/\delta)\xi$ and $\xi\equiv k_{c}\sqrt{\hbar/2M\omega_{a}}
$ is the Lamb-Dicke parameter.

The time evolution
operator takes the form \cite{manko,bose56}
\begin{equation}
\hat{U}(\tau)=e^{ik^{2}\left( \hat{c}^{\dagger}\hat{c}\right) ^{2}\left(
\tau-\sin\tau\right) }e^{k\hat{c}^{\dagger}\hat{c}\left( \eta\hat
{a}^{\dagger}-\eta^{\ast}\hat{a}\right) }e^{-i\tau\hat{a}^{\dagger}\hat{a}
}\,,  \label{Trapped_U2}
\end{equation}
where $\tau\equiv\omega_{a}t$, $k\equiv g/\omega_{a}$, $\eta\equiv1-e^{-i
\tau }$. In writing down Eq.(\ref{Trapped_U2}) we have disregarded
the free radiation field evolution term.

We now consider as initial condition of the ion-cavity system the tensor product
of a coherent state of amplitude $\alpha$ for the cavity mode and a position
squeezed state of amplitude $\beta$ and squeezing parameter $\varepsilon
=r\exp\left\{ 2i\phi\right\} $ for the ion center of mass (see Ref.~\cite{wineRMP}
for the preparation of motional squeezed states of the ion CM)
\begin{equation}
\left| \Psi\left( 0\right) \right\rangle =\left| \beta,\varepsilon
\right\rangle _{a}\otimes\left| \alpha\right\rangle _{c}\,,~~\left|
\beta,\varepsilon\right\rangle _{a}=\hat{D}\left( \beta\right) \hat
{S}\left( \varepsilon\right) \left| 0\right\rangle _{a}\,,
\label{Trapped_CI}
\end{equation}
where $\hat{D}\left( \beta\right)$ is the displacement operator and
$\hat {S}\left( \varepsilon\right) $ is the squeezing operator  \cite{QOptics}.
In order to compute
the time-evolved state $\left| \Psi\left( \tau\right) \right\rangle =\hat{U}
\left( \tau\right) \left| \Psi\left( 0\right) \right\rangle $, we expand the
optical mode coherent state $\left| \alpha\right\rangle _{c}$ in number
states $\left| n\right\rangle _{c}$ and use the Baker-Campbell-Hausdorff
expansion \cite{bch}. After some algebra, we obtain
\begin{align}
\left| \Psi\left( \tau\right) \right\rangle & =\frac{1}{\sqrt{\cosh r}}
\exp\left[{-\frac{\Gamma\beta^{\ast2}+\left| \alpha\right| ^{2}}{2}}  \right]\notag\\
&\times
\sum_{n=0}^{\infty}\frac{\alpha^{n}}{\sqrt{n!}}\exp[\Gamma\beta
^{\ast}e^{-i\tau}\left( \hat{a}^{\dagger}-nk\eta^{\ast}\right)
 -(\Gamma/2)e^{-2i\tau}\left( \hat{a}^{\dagger}-nk\eta^{\ast}\right)
^{2}+ik^{2}n^{2}\left( \tau-\sin\tau\right) ]  \notag \\
& \times\left| \beta e^{-i\tau}+nk\eta\right\rangle _{a}\otimes\left|
n\right\rangle _{c}\,,  \label{statoTemp}
\end{align}
where $\left| \beta e^{-i\tau}+nk\eta\right\rangle _{a}$ are coherent states
of the ion CM motion and
$\Gamma\equiv\exp\left\{ 2i\phi\right\} \tanh r$.
Inserting the identity decomposition in the basis of
ion CM coherent states, $\hat{I}=\int d^{2}\gamma\left| \gamma\right\rangle
_{a}\left\langle \gamma\right| /\pi$ in Eq.~(\ref{statoTemp}), we obtain
\begin{equation}
\left| \Psi\left( \tau\right) \right\rangle =\mathcal{A}\sum_{n=0}^{\infty }
\mathcal{B}_{n}\int\frac{d^{2}\gamma}{\pi}~e^{\mathcal{C}_{n,\gamma}}\left|
\gamma\right\rangle _{a}\otimes\left| n\right\rangle _{c}\,,
\label{Trapped_state1}
\end{equation}
where we have defined
\begin{align}
\mathcal{A} & \equiv\left( \cosh r\right) ^{-1/2}e^{-\frac{\Gamma
\beta^{\ast2}+\left| \alpha\right| ^{2}}{2}},  \label{Trapped_A} \\
\mathcal{B}_{n} & \equiv\frac{\alpha^{n}}{\sqrt{n!}}~e^{ik^{2}n^{2}\left(
\tau-\sin\tau\right) -\Gamma\zeta_{n}^{\ast}(\beta^{\ast}+\zeta_{n}^{\ast
}/2)-\left| \beta+\zeta_{n}\right| ^{2}/2},  \label{Trapped_B}
\end{align}
\begin{equation}
\mathcal{C}_{n,\gamma}\equiv\lbrack\Gamma(\beta^{\ast}+\zeta_{n}^{\ast})+
\beta+\zeta_{n}]e^{-i\tau}\gamma^{\ast}-(\Gamma/2)e^{-2i\tau}\gamma^{\ast
2}-\left| \gamma\right| ^{2}/2,  \label{Trapped_C}
\end{equation}
with
\begin{equation}
\zeta_{n}\equiv nk(e^{i\tau}-1).  \label{Trapped_zita}
\end{equation}

At the (scaled) time $\tau$ we measure the intracavity quadrature $\hat{X}
=(\hat{c}+\hat{c}^{\dagger})/\sqrt{2}$ obtaining the result $X$ \cite{fast}.
As a consequence, the cavity mode is projected onto the
corresponding quadrature eigenstate $\left| X\right\rangle $, while the
ion CM motion is disentangled from the cavity mode and it is projected onto
the state with wave-function
\begin{equation}
\Phi(x)=\frac{N_{X,\tau}}{\pi^{1/4}\sqrt{1-\Gamma e^{-2i\tau}}}\mathcal{A
}\sum_{n=0}^{\infty}\left\langle X\right. \left| n\right\rangle _{c}
\mathcal{B}_{n}e^{\mathcal{D}_{n}(x)}\,,  \label{nearatomicfield}
\end{equation}
where $N_{X,\tau}$ is a normalization constant, and
$x$ now stands for the dimensionless
ion CM\ position, i.e.  $x\rightarrow x\sqrt{M\omega_{a}/\hbar}$. Furthermore
\begin{align}
\mathcal{D}_{n}(x) \equiv\frac{1}{2}(\Gamma-e^{2i\tau})^{-1}\{(
\Gamma+e^{2i\tau})x^{2}+[\Gamma(\beta^{\ast}+\zeta_{n}^{\ast})
 +\beta+\zeta_{n}][\Gamma(\beta^{\ast}+\zeta_{n}^{\ast})+\beta+\zeta _{n}-2
\sqrt{2}e^{i\tau}x]\}\,,  \label{Trapped_D}
\end{align}
with $\langle X|n\rangle_{c}=\pi^{-1/4}\left( 2^{n}n!\right)
^{-1/2}H_{n}(X)e^{-X^{2}/2}$, $H_{n}(X)$ being the $n^{th}$
Hermite polynomial.

\subsection{Homodyning with a zero outcome}

\begin{figure}[ptbh]
\vspace{-0.0cm}
\par
\begin{center}
\includegraphics[width=0.8\textwidth] {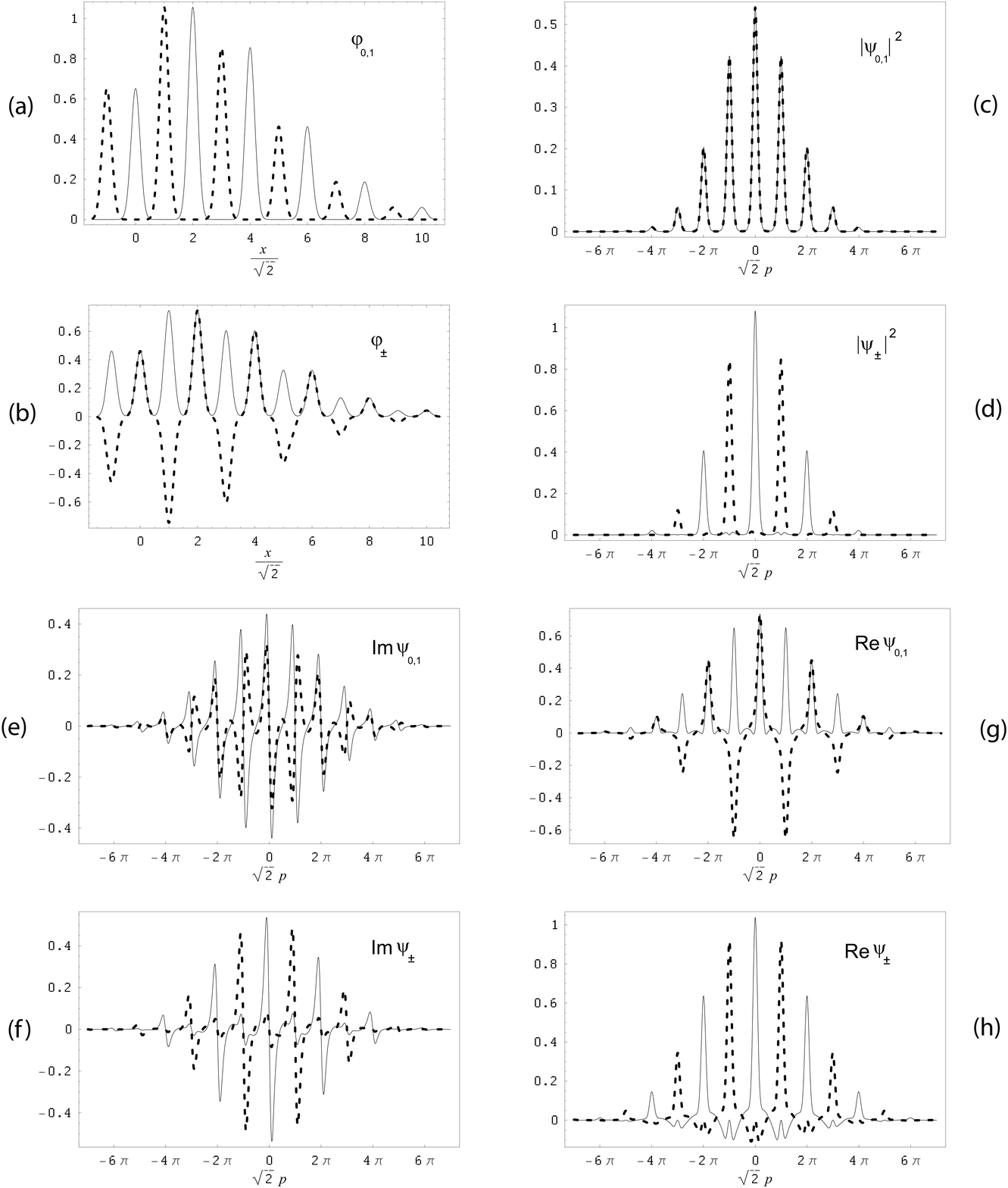}
\end{center}
\par
\vspace{-0.6cm}
\caption{Plot of the wave-functions in space and momentum coordinates of the
approximate codeword states $\widetilde{\left| 0\right\rangle }$, $
\widetilde{\left| 1\right\rangle }$, $\widetilde{\left| +\right\rangle }$
and $\widetilde{\left| -\right\rangle }$ for $\protect\alpha =1.8,$\ $r=1.5$
and $\protect\beta=0$. (a): spatial wave-functions $\protect\varphi_{0}$\ of
$\widetilde{\left| 0\right\rangle }$\ (solid line) and $\protect\varphi_{1}$
\ of $\widetilde{\left| 1\right\rangle }$\ (dashed line) vs $x/\protect\sqrt{
2}$; (b): spatial wave-functions $\protect\varphi_{+}$\ of $\widetilde{
\left| +\right\rangle } $\ (solid line) and $\protect\varphi_{-}$\ of $
\widetilde{\left| -\right\rangle } $\ (dashed line) vs $x/\protect\sqrt{2}$;
(c): momentum probability distributions $\left| \protect\psi_{0}\right| ^{2}
$\ of $\widetilde{\left| 0\right\rangle }$\ (solid line) and $\left| \protect
\psi_{1}\right| ^{2}$\ of $\widetilde{\left| 1\right\rangle }$\ (dashed
line) vs $\protect\sqrt{2}p$; (d): momentum probability distributions $
\left| \protect\psi_{+}\right| ^{2}$\ of $\widetilde{\left| +\right\rangle }$
\ (solid line) and $\left| \protect\psi_{-}\right| ^{2}$\ of $\widetilde{
\left| -\right\rangle }$\ (dashed line) vs $\protect\sqrt{2}p$; (e):
imaginary part of the momentum wave-functions Im$\protect\psi_{0}$\ of $
\widetilde {\left| 0\right\rangle }$\ (solid line) and Im$\protect\psi_{1} $
\ of $\widetilde {\left| 1\right\rangle }$ (dashed line) vs $\protect\sqrt{2}
p$; (f): imaginary part of the momentum wave-functions Im$\protect\psi_{+}$\
of $\widetilde{\left| +\right\rangle }$\ (solid line) and Im$\protect\psi
_{-} $\ of $\widetilde{\left| -\right\rangle }$\ (dashed line) vs $\protect
\sqrt{2}p$; (g): real part of the momentum wave-functions Re$\protect\psi
_{0} $\ of $\widetilde{\left| 0\right\rangle }$\ (solid line) and Re$\protect
\psi_{1}$\ of $\widetilde{\left| 1\right\rangle }$ (dashed line) vs $\protect
\sqrt{2}p$; (h): real part of the momentum wave-functions Re$\protect\psi
_{+} $\ of $\widetilde{\left| +\right\rangle }$\ (solid line) and Re$\protect
\psi_{-}$\ of $\widetilde{\left| -\right\rangle }$\ (dashed line) vs $
\protect\sqrt{2}p$.}
\label{fig09}
\end{figure}

We have now to determine the conditions under which the general ion CM
conditional state of Eq.~(\ref{nearatomicfield}) becomes a comb-like state
which can be taken as approximate codeword state. It is possible to see that
one needs to choose $k=1/2$, that the homodyne measurement has to be
performed at the appropriate time $\tau=\pi$, and that initially the ion has
to be displaced and squeezed in position, which is achieved if we take $
\beta,\varepsilon$ real and positive. To see this we consider a
particular example of homodyne measurement outcome, which makes things
easier to see, i.e. $X=0$. In such a case in fact, the property $
H_{2m}(0)=(-1)^{m}2^{m}(2m-1)!!,$ $H_{2m+1}(0)=0$ $(m=0,1,...)$ allows to
get from~(\ref{nearatomicfield}) the following expression for the ion CM
wave-function
\begin{equation}
\varphi(x)\equiv\left. \Phi(x)\right| _{k=1/2;\tau=\pi;\beta,\varepsilon
\geq0;X=0}=N\sum_{m=0}^{\infty}\nu_{m}\Omega_{m}(x)\,,  \label{fi_x}
\end{equation}
where
\begin{align}
\nu_{m} & \equiv e^{-\alpha^{2}/2}\frac{\alpha^{2m}}{(2m)!!}\,,  \label{ni_m}
\\
\Omega_{m}(x) & \equiv\frac{e^{r/2}}{\sqrt{\pi}}\exp\{-\tfrac{1}{2}e^{2r}[x-
\sqrt{2}(2m-\beta)]^{2}\}\,,  \label{omega_m}
\end{align}
(we have also taken $\alpha$ real and positive) and $N=\left[ \mathcal{P}
(X=0)\right] ^{-1/2}$ with
\begin{equation}
\mathcal{P}(X=0)=\frac{1}{\sqrt{\pi}}\sum_{m,m^{\prime}=0}^{\infty}\nu
_{m}\nu_{m^{\prime}}e^{-2e^{2r}(m-m^{\prime})^{2}}  \label{Prob_0_trapped}
\end{equation}
being the probability density corresponding to the outcome $X=0$ of the
homodyne measurement. The state of Eq.~(\ref{fi_x}) represents, in fact, a
superposition of position squeezed states centered in $x=\sqrt{2}(2m-\beta)$
, $(m=0,1,2...)$, which is however also a superposition of squeezed states
in the momentum space $p$, as one can verify by computing the Fourier
transform
\begin{align}
\psi(p) & =\frac{1}{\sqrt{2\pi}}\int\varphi(x)e^{-ipx}dx \notag \\
& =\frac{N}{\sqrt{\pi}}\exp[-(r+e^{-2r}p^{2})/2]\sum_{m=0}^{\infty}\nu
_{m}e^{-i\sqrt{2}p(2m-\beta)}  \label{psi_p}
\end{align}
(see Figs.~\ref{fig09}(c), \ref{fig09}(e), \ref{fig09}(g)). Therefore, also
in the present scheme, the homodyne measurement of the cavity mode
conditionally generates the desired comb-like state, which we can take as
the approximate codeword state $\widetilde{\left| 0\right\rangle }$. This
state depends upon the three dimensionless parameters $\alpha$, $\beta$, and
$r$. The parameter $r$ is obviously responsible for squeezing in $x$ and
increases the number of spikes in $p$ (as we can see from Eqs.~(\ref{omega_m}),
and (\ref{psi_p})), while $\alpha$ causes squeezing in $p$ and increases
the number of spikes in $x$. This fact can be seen if we consider the
probability to have the $m^{th}$ spike $\Omega_{m}(x)$ in the wave-function $
\varphi(x)$, $P(m)=\nu_{m}/\sum_{m=0}^{\infty}\nu_{m}$. In fact one can see
that the mean value $\bar {m}$ and its standard deviation $\Delta m$ are
approximately given by
\begin{equation}
\bar{m}\simeq\alpha^{2}/2\text{\ , \ }\Delta m\simeq\alpha/\sqrt{2}\,,
\label{momenti}
\end{equation}
implying that the number of spikes in $x$ grows linearly with $\alpha$.
These arguments show that the generated approximate comb-like state
approaches the ideal codeword state $\bar{|0\rangle}$ of Sec.~\ref{ideal}
for $\alpha ,r\rightarrow\infty$ and that we have to take these latter
parameters as large as possible. By varying the parameter $\beta$ instead,
one simply shifts the state along $x$.

Therefore, by comparing with the ideal codeword state of Eq.~(\ref{ideal_0}),
we have that the state of Eq.~(\ref{fi_x}) approximates it (except for an
unimportant shift by $\beta $) with $\theta =\sqrt{2}$. The approximate
codeword $\widetilde{\left| 1\right\rangle }\equiv \hat{D}_{x}(-\sqrt{2})
\widetilde{\left| 0\right\rangle }$ is generated by a position shift
equivalent to a change of the parameter $\beta $, $\Delta \beta =1$, and
which can be again realized by applying a suitable electric field (or laser)
pulse (see Ref.~ \cite{wineRMP}). After seeing how the two basis codeword
states are generated, let us now see how to generate a generic superposition
of the two, $a\widetilde{\left| 0\right\rangle }+b\widetilde{\left|
1\right\rangle }$. These superpositions can be generated using conditional
displacement schemes analogous to those used, for example, in the
manipulation of quantum states of trapped ions \cite{wineRMP} and which
exploit the coupling of a motional degree of freedom with an internal
transition of the ion. Schematically these schemes proceed as follows. The
atom is prepared in the tensor product state $\widetilde{\left|
0\right\rangle }\otimes \left[ a|g\rangle +b|e\rangle \right] $, where $
|e\rangle $ and $|g\rangle $ are two ground state sublevels. Then a laser
pulse which is only coupled to $|e\rangle $ is applied to the atom and its
intensity is tuned so to give exactly a position shift $x\rightarrow x-\sqrt{2}$.
In this way the state of the atom becomes $a|g\rangle \otimes
\widetilde{\left| 0\right\rangle }+b|e\rangle \otimes \widetilde{\left|
1\right\rangle } $. Then a \emph{rf} pulse resonant with the $e\rightarrow g$
transition and transforming $|e\rangle \rightarrow (|e\rangle +|g\rangle )/
\sqrt{2}$ and $|g\rangle \rightarrow (|g\rangle -|e\rangle )/\sqrt{2}$ is
applied, so that the state of the atom becomes $[|g\rangle \otimes (a
\widetilde{\left| 0\right\rangle }+b\widetilde{\left| 1\right\rangle }
)+|e\rangle \otimes (b\widetilde{\left| 1\right\rangle }-a\widetilde{\left|
0\right\rangle })]/\sqrt{2}$. When the internal state of the atom is
measured and it is found equal to $|g\rangle $, the atomic motional state is
conditionally generated in the desired encoded superposition $a\widetilde{
\left| 0\right\rangle }+b\widetilde{\left| 1\right\rangle }$. Particular
examples of these superpositions are the eigenstates of the encoded
phase-flip operator $\bar{X}$, whose approximate versions are given by $
\widetilde{\left| \pm \right\rangle }\equiv \lbrack \widetilde{\left|
0\right\rangle }\pm \widetilde{\left| 1\right\rangle }]/\mathcal{N}_{\pm }$,
where $\mathcal{N}_{\pm }^{2}=2[1\pm \widetilde{\left\langle 0\right. }
\widetilde{\left| 1\right\rangle }]$ ($\widetilde{\left| 0\right\rangle }$
and $\widetilde{\left| 1\right\rangle }$ are not orthogonal in general also
in this scheme). It is interesting to notice that the corresponding
wave-functions of these latter codeword states $\widetilde{\left| \pm
\right\rangle }$ in the momentum space are given by
\begin{align}
\psi _{\pm }\left( p\right) & \equiv \left\langle p\right. \widetilde{\left|
\pm \right\rangle }=\pi ^{-1/2}N\mathcal{N}_{\pm }^{-1}~e^{i\sqrt{2}\beta
p-(r+e^{-2r}p^{2})/2}
  (1\pm e^{i\sqrt{2}p})\sum_{m=0}^{\infty }\nu _{m}e^{-i2\sqrt{2}mp}
\label{psi_p_m}
\end{align}
showing that $\left| \psi _{+}\left( p\right) \right| ^{2}$ has spikes at
$p_{n}=\pi (2n)/\sqrt{2}$ while $\left| \psi _{-}\left( p\right) \right| ^{2}$
has spikes at $p_{n}=\pi (2n+1)/\sqrt{2}$ $(n=0,\pm 1,...)$ (see Fig.~\ref
{fig09}(d)). See Fig.~\ref{fig09}(a)-\ref{fig09}(h) for plots of the spatial
and momentum wave-functions of the various approximate codewords $\widetilde{
\left| 0\right\rangle },\widetilde{\left| 1\right\rangle },\widetilde{\left|
+\right\rangle },\widetilde{\left| -\right\rangle }$ in the case $\beta =0$
and for the particular values $\alpha =1.8$ and $r=1.5$.

\subsection{Intrinsic error probability}

As discussed in Sec.~\ref{ideal}, when approximated codewords are used, one
has additional errors (intrinsic errors). In fact, due to the presence of
the tails of the peaks, the recovery process may lead sometimes to a wrong
codeword. The recovery in the spatial variable is performed by measuring the
operator $(\hat{x}+\beta\sqrt{2})($mod$\sqrt{2})$. An intrinsic error in the
recovery process occurs when, given the state $\varphi_{0}\left( x\right)
\equiv\left\langle x\right. \widetilde{\left| 0\right\rangle }$, the
measurement gives a result within one of the error regions: $R_{n}\equiv
\sqrt{2}\left[ 2n-3/2-\beta,2n-1/2-\beta\right] $, $n=0,1...$. The
corresponding intrinsic error\ probability $P_{x,0}$ is equal to the one, $
P_{x,1}$, which we would obtain starting from the state $\varphi_{1}\left(
x\right) \equiv\left\langle x\right. \widetilde{\left| 1\right\rangle }$ and
considering the complementary error region. So we simply have
\begin{equation}
P_{x}=\sum_{n=0}^{\infty}{\displaystyle\int_{\sqrt{2}(2n-\frac{3}{2}-\beta
)}^{\sqrt{2}(2n-\frac{1}{2}-\beta)}}dx\left| \varphi_{0}\left( x\right)
\right| ^{2}.  \label{probabil_x}
\end{equation}
It is possible to prove that, under the condition $r\gtrsim3/2$, allowing us
to use the asymptotic expansion of the error function $\int_{x}^{+\infty
}dte^{-t^{2}}=(2x)^{-1}e^{-x^{2}}[1-O(x^{-2})]$, we can write the following
upper bound for $P_{x}$:
\begin{equation}
P_{x}\lesssim\frac{N^{2}}{\pi\sqrt{2}}e^{-(r+e^{2r}/2)}[e^{-\alpha^{2}}+
\sum_{m=1}^{\infty}(\nu_{m-1}+\nu_{m})^{2}]\,.  \label{stimaPx}
\end{equation}

The recovery in the momentum space is instead performed by measuring the
operator $\hat{p}($mod$\pi/\sqrt{2})$. An intrinsic error in the recovery
process occurs when, given the state $\psi_{+}\left( p\right) $, the
measurement gives a result within one of the error regions $
R_{n}^{+}\equiv(\pi/\sqrt{2})\left[ 2n+1/2,2n+3/2\right] $ for integer $n$,
or, given the state $\psi_{-}\left( p\right) $, the measurement gives a
result within one of the error regions $R_{n}^{-}\equiv(\pi/\sqrt{2})\left[
2n-1/2,2n+1/2\right] $ for integer $n$. The corresponding intrinsic error
probabilities are given by
\begin{equation}
P_{p,\pm}=\sum_{n=-\infty}^{+\infty}\int_{R_{n}^{\pm}}dp\left| \psi_{\pm
}\left( p\right) \right| ^{2}.  \label{probp_p_m}
\end{equation}
Inserting Eq.~(\ref{psi_p_m}) into Eq.~(\ref{probp_p_m}) we get, after some
algebra,
\begin{equation}
P_{p,\pm}\simeq\frac{2}{\pi}\frac{N^{2}}{\mathcal{N}_{\pm}^{2}}e^{-r}\left[
K_{0,0}^{\pm}\sum_{m=0}^{\infty}\nu_{m}^{2}+2\sum_{m>m^{\prime}=0}^{\infty}
\nu_{m}\nu_{m^{\prime}}K_{m,m^{\prime}}^{\pm}\right] \,,  \label{stimaPp}
\end{equation}
where
\begin{align}
K_{m,m^{\prime}}^{\pm}
\equiv\sum_{n=-\infty}^{+\infty}\int_{R_{n}^{\pm}}dp\ \exp(-e^{-2r}p^{2})
(1\pm\cos\sqrt{2}p)\cos[2\sqrt{2}p(m-m^{\prime})]\text{ }.
\label{Trapped_K}
\end{align}
To estimate the quality of the overall encoding procedure provided by this
scheme, we have to consider a mean intrinsic error probability $\bar{P}_{e}$
, which is obtained, in general, by averaging over all the possible encoded
qubit states. Using the above definitions, we have that the mean intrinsic
error probability $\bar{P}_{e}$ satisfies the inequality
\begin{equation}
\bar{P}_{e}\lesssim\max\{P_{x},P_{p,+},P_{p,-}\}\equiv P_{\max}\,,
\label{pmax}
\end{equation}
which defines the maximum intrinsic error probability $P_{\max}$, providing
therefore a good characterization of the proposed encoding scheme. However,
in the considered physical configuration, in a large and significant region
of parameters, it is $P_{p,+}\simeq P_{p,-}\equiv P_{p}\gg P_{x}$ so that we
can take as upper bound for $\bar{P}_{e}$ simply the quantity $P_{p}$.

Let us therefore study the behavior of this upper bound for the intrinsic
error probability in the case of system parameters corresponding to a
typical experimental cavity QED situation. As we have seen, the amplitude $
\alpha$ of the cavity mode coherent state and the squeezing parameter $r$
the ion CM position are the relevant parameters to study, and they should be
as large as possible. However, as in the preceding scheme, we have derived
our approximate codeword states by making some assumptions and therefore $
\alpha$ and $r$ cannot be freely chosen. A first limitation comes from the
Lamb-Dicke limit. In practice this implies that during all the process of
generation of the comb-like state the ion has to remain localized within a
region smaller than $\lambda_{c}$ so that the linearization of the optical
potential is valid. We can roughly impose $\left| \bar{x}\pm\Delta x\right|
\lesssim\lambda_{c}/8$, where $\bar{x}$ is the ion mean position and $\Delta
x$ is its mean position spread. Using Eqs.~(\ref{fi_x})-(\ref{omega_m})
and~(\ref{momenti}), we can see that the ion comb-like state $\varphi(x)$ is
substantially confined in the interval $\sqrt{2}[2m_{-}-\beta,2m_{+}-\beta]$,
where $m_{\pm}\equiv \max(\alpha^{2}/2\pm\alpha/\sqrt{2},0)$. Using this
fact, we can see that the Lamb-Dicke limit implies
\begin{equation}
(2m_{+}-\beta)\frac{2\xi}{k_{c}}\leq\beta\frac{2\xi}{k_{c}}\leq\frac
{\lambda_{c}}{8}\,,  \label{L_Dicke}
\end{equation}
which is equivalent to require
\begin{equation}
\beta\leq\frac{\pi}{8\xi}\equiv\beta_{\max}\,,~\alpha\leq\frac{\sqrt{
1+4\beta }-1}{\sqrt{2}}\equiv\alpha_{\max}(\beta)\,.  \label{alfamax}
\end{equation}
In order to reach the maximum value of $\alpha$ we choose $
\beta=\beta_{\max} $ so that
\begin{equation}
\alpha\leq\alpha_{\max}(\beta_{\max})=\frac{\sqrt{1+\pi/(2\xi)}-1}{\sqrt{2}}
\equiv\alpha_{\max}~.  \label{Trapped_alfa}
\end{equation}
Therefore we have found an ``optimal'' value $\beta_{max}$ for $\beta$, and
a corresponding upper bound $\alpha_{max}$ for $\alpha$ imposed by the
adoption of the Lamb-Dicke approximation. Moreover the Lamb-Dicke limit has
to be satisfied also by the ion CM state at the beginning, and this implies
a condition on the squeezing parameter $r$ which is roughly given by $r>3/2$.

Finally, the large detuning approximation imposes another limitation on $
\alpha$. In fact, this condition reads $\alpha\lesssim\delta/2g_{0}$ and
using the condition for having a comb-like state $k=1/2$, implying $
\delta=2\xi g_{0}^{2}/\omega_{a}$, we get
\begin{equation}
\alpha\leq\alpha_{\max}^{\prime}\equiv\sqrt{\frac{\hbar}{20M\omega_{a}^{3}}}
g_{0}k_{c}.  \label{Trapped_alfa_2}
\end{equation}
To summarize, the adopted approximations imply the conditions $r\gtrsim3/2$,
$\alpha\leq\alpha_{\max}^{\prime\prime}\equiv\min(\alpha_{\max},\alpha_{\max
}^{\prime})$ and $\beta=\beta_{\max}$.

However, despite these constraints, one can find suitable values, for $
\alpha $ and $r$, such that these bounds are satisfied and a low value for
the estimated intrinsic error probability is achievable.

\begin{figure}[tbph]
\vspace{+0.3cm}
\par
\begin{center}
\includegraphics[width=0.4\textwidth]{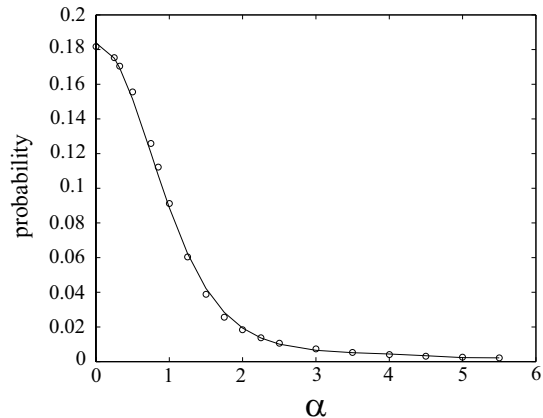}
\end{center}
\par
\vspace{-0.5cm}
\caption{Upper bound $P_{p}$ for the mean error probability $\bar{P}_{e}$ as
a function of $\protect\alpha $ in the interval $0\leq \protect\alpha \leq
5.5$ and for $r=1.5$ (the curve is indistinguishable from those
corresponding to $r=2$ and $r=3$). Numerical data (circles) are fitted by an
exponential curve (line).}
\label{fig08}
\end{figure}

We have done a numerical study of $P_{x},P_{p,+},P_{p,-}$ in the interval $
0\leq\alpha\leq5.5$ and for $r=1.5,2,3$. As anticipated above it is $
P_{x}\ll P_{p}\simeq P_{p,+},P_{p,-}$ so that the intrinsic error
probability is practically determined by $P_{p}$. This upper bound $P_{p}$
weakly depends on $r$ and, in fact, it has the same behavior for $r=1.5,2,3$
, while it has a significant dependence upon $\alpha$. As expected, one has
better results for increasing $\alpha$ (see Fig.~\ref{fig08}), even though
one cannot take too large values for it because of the Lamb-Dicke and large
detuning approximations employed here.

For example, considering the case of a trapped $Ca^{+}$ ion (with $\lambda
_{0}\simeq 866$nm), if we choose experimentally realistic values such as $
\omega _{a}\simeq 400$KHz and $g_{0}\simeq 3.8$MHz, from Eqs.~(\ref
{Trapped_alfa}) and~(\ref{Trapped_alfa_2}) one gets $\alpha _{\max }^{\prime
\prime }\simeq 1$ corresponding to $\beta =\beta _{\max }\simeq 1.2$. In
correspondence of these values for $\alpha $ and $\beta $, and choosing $
r=1.5$ one has a mean intrinsic error probability $\bar{P}_{e}\lesssim 9\%$,
as we can see from Fig.~\ref{fig08}. These particular values of $\alpha $
and $r$ imply a \emph{success probability} $\mathcal{P}(X=0)\simeq 26\%$
according to Eq.~(\ref{Prob_0_trapped}). Note that, thanks to its
exponential-decaying behavior, the value of the intrisic error probability
can rapidly be improved by accessing stronger ponderomotive interactions
(i.e., higher $g_{0}-$couplings).

\section{Conclusion\label{conclusion}}

For any quantum information processing to become a reality the task of
providing adequate error correction needs to be fulfilled. As the quantum
mechanical oscillator is a simply and prevalent model in the study of
quantum mechanics, it appears to be a natural test bed for such purposes.
Thus we have discussed how to embed a qubit in a continuous quantum system,
so that its redundancy can be used to correct errors which arise from
unwanted interactions with the environment. In particular we have shown that
ponderomotive interaction suffices to this end and we have proposed a
physical scheme in which a trapped ion within a cavity is considered.

In this configuration, we are able to show that sufficiently low values of
the intrinsic error probability are reachable so that idealized
states \cite{preskill} can be effectively engineered with current technology.
For the sake of simplicity, throughout the paper we
have considered protocols conditioned to a given result of a homodyne
measurement of the intracavity field (the case corresponding to the outcome $
X=0$ to be specific). However in practical situations one has to
consider a finite interval of acceptable measurement results. Then, it is
possible to see that the success probability is improved by enlarging the
interval at expenses of an increased error probability.

\end{document}